Title: State Complexity of Reversible Watson-Crick Automata

Authors: Kingshuk Chatterjee[1], Debayan Ganguly (corresponding author)[2], Kumar Sankar Ray[3]

Affiliations: [1]Government College of Engineering and Ceramic technology, Kolkata-700010
[2]Government College of Engineering and Leather technology, Kolkata-700106
[3]Electronics and Communication Sciences Unit, Indian Statistical Institute, Kolkata-700108.

# State compexity of Reversible Watson-Crick automata

Kingshuk Chatterjee[1], Debayan Ganguly[2], Kumar Sankar Ray[3]

[1]Government College of Engineering and Ceramic technology, Kolkata-700010
[2]Government College of Engineering and Leather technology, Kolkata-700106
[3]Electronics and Communication Sciences Unit, Indian Statistical Institute, Kolkata-700108.
[1] kingshukchaterjee@gmail.com, [2] debayan3737@gmail.com, [3] ksray@isical.ac.in

*Abstract: Reversible Watson-Crick automata introduced by Chatterjee et.al. is a reversible variant of an Watson-Crick automata. It has already been shown that the addition of DNA properties to reversible automata significantly increases the computational power of the model. In this paper, we analyze the state complexity of Reversible Watson-Crick automata with respect to non-deterministic finite automata. We show that Reversible Watson-Crick automata in spite of being reversible in nature enjoy state complexity advantage over non deterministic finite automata. The result is interesting because conversion from non deterministic to deterministic automata results in exponential blow up of the number of states and classically increase in number of heads of the automata cannot compensate for non-determinism in deterministic and reversible models.*

## I. Introduction

Interest in Reversible automata started when Bennet[1] showed that Reversible Turing machine has the same computational power as a non-deterministic Turing machine. Researchers started exploring restricted versions of reversible automata Pin[2] showed that languages accepted by reversible automata is strictly a proper subset of regular languages. Morita[3] explored the two way variant of multi-head reversible automata and showed that it has the same computational power as two way multi-head deterministic finite automata. Kutrib et.al.[4] showed that one way multi-head reversible finite automata is strictly weaker than its deterministic variant. Watson-Crick automata [5] are finite automata having two independent heads working on double strands where the characters on the corresponding positions of the two strands are connected by a complementarity relation similar to the Watson-Crick complementarity relation. The movement of the heads although independent of each other is controlled by a single state. Chatterjee et.al.[6] introduced a reversible variant of Watson-Crick automata and showed that it can accept all regular languages. Moreover, reversible Watson-Crick automata were able to accept languages which are not accepted by multi-head deterministic finite automata. State and transition complexity of Watson-Crick automata are discussed in detail in[7]. In this paper, we compare the state complexity of Reversible Watson-Crick automata with respect to non-deterministic finite automata.

The main claims of this paper are as follows:

1. For every non deterministic finite automaton accepting a language L using n states there exist a reversible Watson-Crick automata which accepts the same language L using n+2 states.

2. Non-deterministic finite automaton requires at least $2^k+1$ states to accept the language $L_k=\{x^n| n\geq 1, x\in \{a,b\}^k\}$ whereas reversible Watson-Crick automata requires $2k+2$ states to accept $L_k$.

## II. Basic Terminology

The symbol V denotes a finite alphabet. The set of all finite words over V is denoted by $V^*$, which includes the empty word $\lambda$. The set of all non-empty words over the alphabet V is denoted by $V^+=V^*-\{\lambda\}$. For $w \in V^*$, the length of w is denoted by $|w|$. Let $u \in V^*$ and $v \in V^*$ be two words and if there is some word $x \in V^*$, such that v=ux, then u is a prefix of v, denoted by $u \leq v$. Two words, u and v are prefix comparable denoted by $u\sim_p v$ if u is a prefix of v or vice versa.

**Reversible Watson-Crick automata**

A one-way reversible Watson-Crick automaton (Rev-WKA) is a system M= (Q, V, #, $, $q_0$, F, ρ, δ) where

1) Q is the finite set of states.

2) V is the finite set of input symbols.

3) #∉V is the left end marker for both upper and lower strand.

4) $∉V is the right end marker for both upper and lower strand.

5) $q_0 \in Q$ is the initial state.

6) F⊆Q is the set of accepting states

7) ρ is the symmetric complementarity relation i.e. ρ:V→V where ρ is defined for every x∈V and ρ can be non-injective and

8) $\delta: Q \times (V \cup \{\#,\$\})^2 \to Q \times \{0,1\}^2$ is the partial transition function where 1 means to move the head one position to the right and 0 means to keep the head in the current position. Whenever $\delta(q',(a_1,a_2))=(q,(d_1,d_2))$ is defined then $d_i=0$ if $a_i=\$$ $1 \leq i \leq 2$.

To make the automaton reversible the following two restrictions are imposed on the partial transition function δ;

1) for any two transitions $\delta(q',(a_1,a_2))=(q,(d_1,d_2))$ and $\delta(q'',(a_1',a_2'))=(q,(d_1',d_2'))$ it holds that $d_1'=d_1$ and $d_2'=d_2$.

2) for any two transitions $\delta(q',(a_1,a_2))=(q,(d_1,d_2))$ and $\delta(q'',(a_1',a_2'))=(q,(d_1,d_2))$ it holds that either $a_1 \neq a_1'$ or $a_2 \neq a_2'$.

Condition (1) ensures transitions resulting in the same state have to move the heads in the same way. Whereas condition (2) (termed by Morita[3] as the reversibility condition) ensures forward as well as backward determinism in the automaton. In definition 3, we have replaced Kutrib et. al.'s reachability condition [4] for reversible automaton by Morita's reversibility condition.

The configuration of a Rev-WKA M= (Q, V, #, $, $q_0$, F, ρ, δ) at some time t≥0 is a quadruple $c_t=(w_1,w_2,q,P)$ where $w_1 \in V^*$ and $w_2 \in V^*$ where $w_2$ (the string in the second strand) is determined in the following manner:

If $w_1=\lambda$ then $w_2=\lambda$, if $w_1=x_1x_2...x_n$ where $|w_1|=n$ and $x_i \in V$, $1 \leq i \leq n$ then $w_2=\rho(x_1)\rho(x_2),...,\rho(x_n)$. The symbol $q \in Q$ is the current state and $P=(p_1, p_2) \in \{0,1,...,|w_1|+1\}^2$ gives the current head position. If $p_1=0$, then the head on the upper strand is scanning the symbol # on the upper strand. If it satisfies $1 \leq p_1 \leq |w_1|$, then the upper head is scanning the $p_1$th symbol of $w_1$ in the upper strand and if $p_1=|w_1|+1$ then the upper head is scanning the end marker symbol $ at the end of $w_1$. The interpretation of the values of $p_2$ is similar to that of $p_1$. The only difference being, $p_2$ denotes the position of the lower head placed on the lower strand and the lower head reads $w_2$. The initial configuration for input is set to $(w_1,w_2,q_0,(0,0))$ where string in the upper strand is of the form $\#w_1\$$ and the string in the lower strand is of the form $\#w_2\$$. During its course of computation M goes through a sequence of configurations. One step from a configuration to its successor configuration is denoted by ⊢. Let $w_1=c_1c_2...c_n$, $c_0=\#$ and $c_{n+1}=\$$, $w_2$ is obtained from $w_1$ in a manner as described above and let it be of the form $w_2=b_1b_2,...,b_n$, $b_0=\#$ and $b_{n+1}=\$$. We write $(w_1,w_2,q,(p_1,p_2)) \vdash (w_1,w_2,q',(p_1+d_1,p_2+d_2))$ if and only if $\delta(q,(c_{p_1},b_{p_2}))=(q',(d_1,d_2))$ exists. The reflexive and transitive closure of ⊢ is denoted by $\vdash^*$.

The language accepted by Rev-WKA in the upper strand is precisely the set of words $w_1$ such that there is some computation beginning with $\#w_1\$$ in the upper strand and $\#w_2\$$ in the lower strand. The string $w_2$ is obtained in the same manner as described in the beginning of this section and Rev-WKA halts in an accepting state. A Rev-WKA halts when the transition function is not defined for the current situation.

$L(M)=\{w_1 \in V^* | (w_1,w_2,q_0,(0,0)) \vdash^* (w_1,w_2,q,(p_1,p_2))$, $w_2=\lambda$ if $w_1=\lambda$ otherwise $w_2= \rho(x_1)\rho(x_2),...,\rho(x_n)$ if $w_1=x_1x_2,...,x_n$ where $x_i \in V$, $1 \leq i \leq n$, $q \in F$, and M halts in $(w_1,w_2,q,(p_1,p_2))\}$.

An input in the upper strand is accepted if and only if M halts in an accepting state; in all other cases it is rejected. That is if the computation halts in a rejecting state or if the computation runs into an infinite loop. In the case of infinite loop eventually all heads are stationary since the machine described is one-way.

A reversible Watson-Crick automaton is called a ***strongly reversible Watson-Crick automaton*** if the complementarity relation is identity in that particular automaton.

The above definition of Reversible Watson-Crick automaton is obtained from [6].

### III. STATE COMPLEXITY OF REVERSIBLE WATSON-CRICK FINITE AUTOMATA

In this section we compare the state complexity of non-deterministic finite automata with Reversible Watson-Crick automata. We show that Reversible Watson-Crick automata enjoy state complexity advantage over non-deterministic finite automata even though it is reversible.

**Theorem 1:** For every non-deterministic finite automaton which accepts a language L using n states, we can find a reversible Watson-Crick automaton with non-injective complementarity relation which accepts the same language L with n+2 states.

**Proof:** The proof of the above Theorem is similar to the proof in [6] for deterministic finite automata and has been slightly modified to consider state complexity of the constructed reversible Watson-Crick automata and is in two parts. In the first part given a non-deterministic finite automaton M which accepts a language L, we give a construction to obtain a reversible Watson-Crick automaton M' from M and in the second part we show that M' accepts the same language as M.

**First Part**: Given a non-deterministic finite automaton M= (Q,V,$q_0$,F,δ). We construct a reversible Watson-Crick automaton M'= (Q', V', #, $, $q_0'$, F', ρ, δ') from M in the following manner:

We list all the transitions in M in a particular order. Each transition is assigned a symbol '$t_i$' where '$t_i$'∉V and 'i' is the position of the transition in the list. If there are m transitions in M then $V'=V\cup\{t_i|1\leq i\leq m\}$ and complementarity relation $\rho=\{(x,t_1),(t_1,x),...,(x,t_i),(t_i,x),..., (x,t_n),(t_n,x)| \forall x\in V\}$. For each transition $q'\in\delta(q,x)$ in $\delta$ having symbol '$t_i$' assigned to it, we introduce the transition $\delta'(q,(x,t_i))=(q',(1,1))$ in $\delta'$.

Moreover the following transitions are also added to $\delta'$.

1) $\delta'(q_0',(\#,\#))=(q_0,(1,1))$

2) $\delta'(q,(\$,\$))=(q_f,(0,0))$ for all $q\in F$

The set of states of M' i.e. $Q'=Q\cup\{q_0',q_f\}$. If M has n states, M' has n+2 states.

The set of final states of M' i.e. $F'=\{q_f\}$.

The start state of M' is $q_0'$.

**Second Part:** In this part we show that M' constructed from M accepts the same language L.

If M accepts w then there is a sequence of transitions of length |w| that takes M to a final state after consuming w. The complementarity relation $\rho$ of M' relates each symbol to all the transitions in $\delta$, therefore the set of possible complementarity strings for w comprises of all sequences of transitions of length |w|. As M accepts w among the set of possible complementarity strings for w, we will find a complementarity string w' which resemble the sequence of transitions that takes M to a final state after consuming w. Now M' armed with this correct sequence of transitions can deterministically and reversibly (as each transition $q'\in\delta(q,x)$ of M is assigned a unique number $\delta'(q,(x,t_i))=(q',(1,1))$ in M') decide which transitions to take and following this sequence of transitions M' reaches the final state of M, and both its heads are on $. Then the transition $\delta'(q,(\$,\$))=(q_f,(0,0))$ for all $q\in F$ takes M' to its final state after completely consuming its input and thus M' also accepts w.

If M does not accept w then there is no sequence of transitions that takes M to a final state after consumption of w. Thus, no matter what the complementarity string w' of w is, M' simulating M based on w' will never be in a position where it is in a final state of M and the two heads of M' are on $. Thus the transitions of the form $\delta'(q,(\$,\$))=(q_f,(0,0))$ for all $q\in F$ cannot be applied to M', so M' will never enter its final state $q_f$ as a result M' also rejects w.

**Example 1:** Consider the language $L_2=\{x^n| n\geq 1, x\in\{a,b\}^2\}$, reversible Watson-Crick automaton M= (Q, V, #, $, $q_0$, F, $\rho$, $\delta$) can accept the language $L_2$ using 2(2+1)=6 states.

$Q=\{q_0,q_1,q_3,q_4,q_5,q_6\}$, $V=\{a,b,a_1,b_1\}$, $\rho=\{(a,a),(a,a_1),(a_1,a),(b,b),(b,b_1),(b_1,b)\}$, $F=\{q_6\}$. The transitions $\delta$ of M are as follows:

$\delta(q_0,(\#,\#))=(q_0,(1,1))$, $\delta(q_0,(x,x))=(q_1,(0,1))$ $x\in\{a,b\}$, $\delta(q_1,(x,x))=(q_2,(0,1))$ $x\in\{a,b\}$, $\delta(q_2,(a,a_1))=(q_3,(1,1))$, $\delta(q_2,(b,b_1))=(q_3,(1,1))$, $\delta(q_3,(x,x))=(q_4,(1,1))$ $x\in\{a,b\}$, $\delta(q_4,(x,x))=(q_5,(1,1))$ $x\in\{a,b\}$, $\delta(q_5,(x,x))=(q_3,(1,1))$ $x\in\{a,b\}$, $\delta(q_5,(x,\$))=(q_6,(1,0))$ $x\in\{a,b\}$

The above automaton works in the following manner:

After reading the beginning marker in both the tapes, the lower head of M, first shifts 2 places, to the right whereas the upper head remains in its position. Now for every position of the upper head we match with the lower head to see whether they are reading the same values or not, we repeat this process for 2 positions as length of x is 2, if there is a mismatch we reject the input string as there is no transition defined when there is a mismatch and the automaton halts in an non-final state. If there is no mismatch then we repeat this procedure of checking lower head and upper head values repeatedly for 2 times by entering state $q_3$. If the lower head reaches $ then we accept the string by going to state $q_6$ and halting. To ensure reversibility of our model when transitioning from state $q_2$ to $q_3$ we use the transitions $\delta(q_2,(a,a_1))=(q_3,(1,1))$, $\delta(q_2,(b,b_1))=(q_3,(1,1))$ instead of $\delta(q_2,(a,a))=(q_3,(1,1))$, $\delta(q_2,(b,b))=(q_3,(1,1))$ so that when the loop again comes back to state $q_3$ using the transition $\delta(q_5,(x,x))=(q_3,(1,1))$ $x\in\{a,b\}$ from $q_5$ there is no conflict in reversibility condition.

**Theorem 2:** Consider the language $L_k=\{x^n| n\geq 1, x\in\{a,b\}^k\}$, reversible Watson-Crick automaton M= (Q, V, #, $, $q_0$, F, $\rho$, $\delta$) can accept the language $L_k$ using 2(k+1) =2k+2 states.

**Proof:** This proof is a generalisation of the construction used in Example 1, in Example 1, k was equal to 2. The reversible Watson-Crick automaton M= (Q, V, #, $, $q_0$, F, $\rho$, $\delta$) is $Q=\{q_0,q_1,...,q_{2(k+1)}\}$, $V=\{a,b,a_1,b_1\}$, $\rho=\{(a,a),(a,a_1),(a_1,a),(b,b),(b,b_1),(b_1,b)\}$, $F=\{q_{2k+2}\}$.

The transitions $\delta$ of M are as follows:

$\delta(q_0,(\#,\#))=(q_0,(1,1))$, $\delta(q_m,(x,x))=(q_{m+1},(0,1))$ $x\in\{a,b\}$ where m=0,1,2,...,(k-1), $\delta(q_k,(a,a_1))=(q_{k+1},(1,1))$, $\delta(q_k,(b,b_1))=(q_{k+1},(1,1))$, $\delta(q_m,(x,x))=(q_{m+1},(1,1))$ $x\in\{a,b\}$ where m=(k+1),(k+2),...,2k, $\delta(q_{2k+1},(x,x))=(q_{k+1},(1,1))$ $x\in\{a,b\}$, $\delta(q_{2k+1},(x,\$))=(q_{2k+2},(1,0))$ $x\in\{a,b\}$

The above automaton works in the following manner:

After reading the beginning marker in both the tapes using state $q_0$ and transition $\delta(q_0,(\#,\#))=(q_0,(1,1))$, the lower head of M, first shifts k places to the right whereas the upper head remains in its position using the states $q_0,q_1,\ldots,q_k$ and the transitions $\delta(q_m,(x,x))=(q_{m+1},(0,1))$ x$\in$ {a, b} where m=0,1,2,…,(k-1). Now for every position of the upper head we match with the lower head to see whether they are reading the same values or not, we repeat this process for k positions as length of x is l using states $q_{k+1},q_{k+2},\ldots,q_{2k+1}$, if there is a mismatch we reject the input string as there is no transition defined when there is a mismatch and the automaton halts in an non-final state. If there is no mismatch then we repeat this procedure of checking lower head and upper head values repeatedly for k times by entering state $q_{k+1}$ again and again and repeating the above procedure. If the lower head reaches $ then we accept the string by going to state $q_{2k+2}$ and halting. To ensure reversibility of our model when transitioning from state $q_k$ to $q_{k+1}$ we use the transitions $\delta(q_k,(a,a_1))=(q_{k+1},(1,1))$, $\delta(q_k,(b,b_1))=(q_{k+1},(1,1))$ instead of $\delta(q_k,(a,a))=(q_{k+1},(1,1))$, $\delta(q_k,(b,b))=(q_{k+1},(1,1))$ so that when the loop again comes back to state $q_3$ using the transition $\delta(q_{2k+1},(x,x))=(q_{k+1},(1,1))$ x$\in$ {a, b} from $q_{2k+1}$ there is no conflict in reversibility condition. Thus the above automaton accepts $L_k$ using just 2k+2 states.

**Lemma 2:** Number states required for non deterministic finite automata to accept $L_k=\{x^n| n\geq 1, x\in\{a,b\}^k\}$ is greater than $2^k+1$.

**Proof:** The proof of this lemma is in [7].

**Theorem 3:** Reversible Watson-Crick automata has state complexity advantage over non-deterministic finite automata

**Proof:** From, Theorem 2 and Lemma 2 we see that to accept $L_k=\{x^n| n\geq 1, x\in\{a,b\}^k\}$ a non-deterministic finite automaton requires at least $2^k+1$ states whereas Reversible Watson-Crick automaton accepts it using 2k+2 states. Moreover, from Theorem 1 we see no matter the language accepted by non-deterministic finite automaton, Reversible Watson-Crick automaton always accepts it using just two more states than the non-deterministic version, so there is just a difference of a constant number of states, whereas for some languages such as $L_k$ the difference in number of states is exponential to linear. Thus, we can say that Reversible Watson-Crick automata have state complexity advantage over non-deterministic finite automata.

## IV. CONCLUSIONS

In this paper we have compared the state complexity of Reversible Watson-Crick automata with non-deterministic finite automata. We have shown for every non deterministic finite automaton accepting a language L using n states there exist a reversible Watson-Crick automaton which accepts the same language L using n+2 states. Moreover we also showed Reversible Watson-Crick automaton requires 2k+2 states to accept $L_k=\{x^n| n\geq 1, x\in\{a,b\}^k\}$. We already know that a non-deterministic finite automaton requires at least $2^k+1$ states to accept $L_k$. Thus we, concluded that Reversible Watson-Crick automata has state complexity advantage over non-deterministic finite automata.

## REFERENCES

[1] C. H. Bennett, "Logical reversibility of computation," *IBM Journal of Research and Development*, vol. 17, pp. 525–532, nov 1973.
[2] J. E. Pin, "On the languages accepted by finite reversible automata," in *Automata, Languages and Programming*, pp. 237–249, Springer Berlin Heidelberg, 1987.
[3] K. Morita, "Two-way reversible multi-head finite automata," *Fundam. Inf.*, vol. 110, no. 1-4, pp. 241-254, 2011.
[4] M. Kutrib and A. Malcher, "One-way reversible multi-head finite automata," *Theoretical Computer Science*, vol. 682, pp. 149–164, jun 2017.
[5] R. Freund, G. Păun, G. Rozenberg, and A. Salomaa, "Watson-crick finite automata," in *DNA Based Computers III*, pp. 297–327, American Mathematical Society, apr 1999.
[6] K. Chatterjee and K. S. Ray, "Reversible watson-crick automata," *Acta Inf.*, vol. 54, no. 5, pp. 487–499, 2017.
[7] A. Păun and M. Păun, "State and transition complexity of watson-crick finite automata," in *Fundamentals of Computation Theory*, pp. 409–420, Springer Berlin Heidelberg, 1999.